\documentclass[preprint,aps,showpacs,preprintnumbers,amsmath,amssymb]{revtex4}  

\usepackage{graphicx}
\usepackage{dcolumn}
\usepackage{bm}

\begin{document}

\title{Enhancement of Intermediate-Field Two-Photon Absorption by Rationally-Shaped Femtosecond Pulses}
\author{Lev Chuntonov, Leonid Rybak, Andrey Gandman,}
\author{Zohar Amitay}%
\email{amitayz@tx.technion.ac.il} %
\affiliation{Schulich Faculty of Chemistry, Technion - Israel Institute of Technology, Haifa 32000, Israel}

\begin{abstract}
We extend the powerful frequency-domain analysis of femtosecond
two-photon absorption to the intermediate-field regime, which involves
both two- and four-photon transitions. Consequently, we find a broad
family of shaped pulses that enhance the absorption over the
transform-limited pulse. It includes any spectral phase that is
anti-symmetric around half the transition frequency. The spectrum is
asymmetric around it. The theoretical framework and results for Na
are verified experimentally. This work opens the door for rational
femtosecond coherent control in a regime of considerable absorption yields.
\end{abstract}

\maketitle

Following from their coherence over a broad spectrum, shaped
femtosecond pulses allow to coherently control quantum dynamics in
ways that other means cannot
\cite{tannor_kosloff_rice_coh_cont,shapiro_brumer_coh_cont_book,warren_rabitz_coh_cont,rabitz_vivie_motzkus_kompa_coh_cont,dantus_exp_review1_2}.
Among the processes, over which such control has been most effective, are
the important
multiphoton absorption processes 
\cite{dantus_exp_review1_2,silberberg_2ph_nonres1_2,dantus_2ph_nonres_molec1_2,
baumert_2ph_nonres,silberberg_2ph_1plus1,girard_2ph_1plus1,becker_2ph_1plus1_theo,silberberg_antiStokes_Raman_spect,gersh_murnane_kapteyn_Raman_spect,leone_res_nonres_raman_control,
amitay_3ph_2plus1,silberberg-2ph-strong-field,weinacht-2ph-strong-theo-exp,wollenhaupt-baumert1_2,hosseini_theo_multiphoton_strong}.
The control is achieved by shaping the pulse to manipulate the interferences between 
various initial-to-final state-to-state multiphoton pathways.
Constructive interferences enhance the absorption, while destructive
interferences attenuate it.
%
%
Thus, to fully utilize coherent control,
the pulse shaping should ideally be based on identifying first the pathways and their interference mechanism.
When such identification is not feasible, a practical partial
solution is to employ automatic "black-box" experimental
optimization of the pulse shape using learning algorithms
\cite{rabitz_feedback_learning_idea}.
For multiphoton 
processes, the past control studies
\cite{dantus_exp_review1_2,silberberg_2ph_nonres1_2,dantus_2ph_nonres_molec1_2,
baumert_2ph_nonres,silberberg_2ph_1plus1,girard_2ph_1plus1,becker_2ph_1plus1_theo,
silberberg_antiStokes_Raman_spect,gersh_murnane_kapteyn_Raman_spect,leone_res_nonres_raman_control,amitay_3ph_2plus1}
have shown that this ideal line of action is feasible and very powerful once the
photo-excitation picture is available in the frequency domain.
However, so far the frequency domain has been exploited only
in the weak-field regime \cite{dantus_exp_review1_2,silberberg_2ph_nonres1_2,dantus_2ph_nonres_molec1_2,
baumert_2ph_nonres,silberberg_2ph_1plus1,girard_2ph_1plus1,becker_2ph_1plus1_theo,
silberberg_antiStokes_Raman_spect,gersh_murnane_kapteyn_Raman_spect,leone_res_nonres_raman_control,amitay_3ph_2plus1}
associated with low absorption yields (typically up to $\sim$0.1\% population transfer).
There, the N-photon absorption is 
described by
time-dependent perturbation theory of the
lowest non-vanishing order (the N$^{th}$ order)
and, thus, can be 
transformed to the frequency domain.

The present work extends the powerful frequency-domain picture to a regime of sizeable absorption yields,
exceeding the weak-field yields by two orders of magnitude (and more).
In this regime the interfering pathways of N-photon absorption are
the weak-field pathways of N absorbed photons as well as 
additional pathways of M absorbed photons and M$-$N emitted photons (M$>$N).
The corresponding picture is obtained by extending the perturbative
analysis to include a finite number of non-vanishing orders beyond
the lowest one.
We refer to this regime as the intermediate-field regime. It is
distinguished from the strong-field regime where no perturbative
description is valid, which is the one all the past multiphoton
control studies beyond the weak-field regime have focused on
\cite{silberberg-2ph-strong-field,weinacht-2ph-strong-theo-exp,wollenhaupt-baumert1_2,hosseini_theo_multiphoton_strong}.
Specifically, based on 4$^{th}$-order perturbation theory, we
develop here an intermediate-field frequency-domain theoretical
framework to the process of atomic femtosecond two-photon
absorption, which is non-resonant in the weak-field regime (of
2$^{nd}$ order). As a 4$^{th}$-order approach, it involves both the
2$^{nd}$ and 4$^{th}$ perturbative orders associated, respectively,
with two- and four-photon pathways.
Consequently, we find an extensive family of shaped pulses that in
this regime enhance the absorption as compared to the unshaped
transform-limited (TL) pulse.
This is impossible in the weak-field regime
\cite{silberberg_2ph_nonres1_2}.
The theoretical framework and results are verified experimentally.
The approach developed here is general and can be employed in other
cases.

\begin{figure} [htbp]
\includegraphics[scale=0.5]{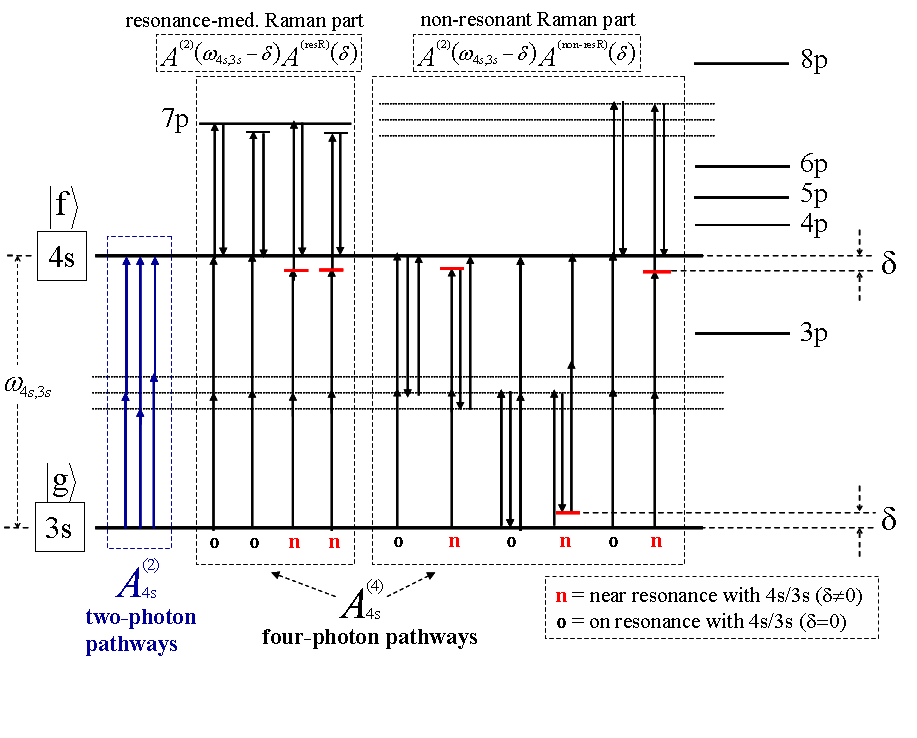} %
\caption{(Color online) Excitation scheme of the intermediate-field
two-photon absorption in Na (not to scale), with representative
two-photon and four-photon pathways from $\left| g \right\rangle
\equiv 3s$
to $\left| f \right\rangle \equiv 4s$. 
Each of the four-photon pathways is either on- or near-resonance with $4s$ or $3s$.
Its Raman part is non-resonant due to the $np$ states ($n$$\ne$7) or resonance-mediated via $7p$.
}
\label{fig_1}
\end{figure}

The physical model system of the study is the sodium (Na) atom (see Fig.~\ref{fig_1}).
Theoretically, we consider an atomic two-photon absorption process
from an initial ground state $\left|g\right>$ to a final excited
state $\left|f\right>$, which are coupled via a manifold of
intermediate states $\left|n\right>$ having the proper symmetry.
The spectrum of the pulse is such that all the
$\left|g\right>$-$\left|n\right>$ and
$\left|f\right>$-$\left|n\right>$ couplings are off-resonant, %
i.e., the spectral amplitude at all the corresponding transition
frequencies is zero: $\left|E(\omega_{gn})\right| =
\left|E(\omega_{fn})\right| = 0$, except for the
$\left|f\right>$-$\left|n_{r}\right>$ resonant coupling for which
$\left|E(\omega_{fn_{r}})\right| \ne 0$.
In the present intermediate-field regime, the final (complex)
amplitude $A_{f}$ of state $\left|f\right>$, following irradiation
with a (shaped) temporal electric field $\varepsilon(t)$, can be
validly described by 4$^{th}$-order time-dependent perturbation
theory. So, in general, it includes non-vanishing contributions from
both the 2$^{nd}$ and 4$^{th}$ perturbative orders:
\begin{equation}
A_{f} = A_{f}^{(2)} + A_{f}^{(4)} \; .
\label{eq:total-amp}
\end{equation}
Within the frequency-domain framework,
the spectral field 
$E(\omega) \equiv \left|E(\omega)\right| \exp \left[ i\Phi(\omega)
\right]$ is given as the Fourier transform of $\varepsilon(t)$, with
$\left|E(\omega)\right|$ and $\Phi(\omega)$ being the spectral
amplitude and phase at frequency $\omega$.
For the TL pulse: $\Phi(\omega)=0$. 
We also introduce the normalized spectral field
$\widetilde{E}(\omega) \equiv E(\omega) / \left|E_{0}\right|$
representing the pulse shape, with $\left|E_{0}\right|$ being the
maximal spectral amplitude.
As shown for the weak-field regime \cite{silberberg_2ph_nonres1_2},
the 2$^{nd}$-order term $A_{f}^{(2)}$
interferes all the $\left|g\right>$-$\left|f\right>$ pathways of two
absorbed photons (see examples in Fig.~\ref{fig_1}) and is given by
\begin{equation}
A_{f}^{(2)} =
- \frac{1}{i \hbar^{2}} \left|E_{0}\right|^{2} A^{(2)}(\omega_{fg}) 
\label{eq1:amp-2nd-order}
\end{equation}
\begin{equation}
A^{(2)}(\Omega) = \mu_{fg}^{2}
\int_{-\infty}^{\infty}\widetilde{E}(\omega)\widetilde{E}(\Omega-\omega)d\omega \; , %
\label{eq2:amp-2nd-order} %
\end{equation}
where $\omega_{fg}$ and $\mu_{fg}^{2}$ are the
$\left|g\right>$-$\left|f\right>$ transition frequency and 
effective non-resonant two-photon coupling.
The TL pulse induces fully constructive interferences within
$A_{f}^{(2)}$ and, thus, the maximal $\left|A_{f}^{(2)}\right|$ and
maximal weak-field non-resonant two-photon absorption
\cite{silberberg_2ph_nonres1_2}.
%
The 4$^{th}$-order term $A_{f}^{(4)}$ is 
more complicated than $A_{f}^{(2)}$  %
and we have analytically calculated it to be given by               
\begin{eqnarray}
A_{f}^{(4)}  & = & - \frac{1}{i \hbar^{4}} \left|E_{0}\right|^{4}
\left[ i \pi A^{(2)}(\omega_{fg}) A^{(R)}(0) - \wp
\int_{-\infty}^{\infty} d\delta \frac{1}{\delta}
A^{(2)}(\omega_{fg}-\delta) A^{(R)}(\delta) \right] ,
\label{eq1:amp-4th-order}
\end{eqnarray}
where $A^{(2)}(\Omega)$ is defined above, and 
\begin{eqnarray}
A^{(R)}(\Delta\Omega) & = & A^{(\scriptsize{\textrm{non-res}}R)}(\Delta\Omega) + A^{(\scriptsize{\textrm{res}}R)}(\Delta\Omega)  %
\label{eq2:amp-4th-order}
\\
A^{(\scriptsize{\textrm{non-res}}R)}(\Delta\Omega) & = &
(\mu_{ff}^{2}+\mu_{gg}^{2}) \int_{-\infty}^{\infty} \widetilde{E}(\omega + \Delta\Omega) \widetilde{E}^{*}(\omega) d\omega   %
\label{eq3:amp-4th-order}
\\
A^{(\scriptsize{\textrm{res}}R)}(\Delta\Omega) & = & |\mu_{fr}|^{2}
\left[ i \pi
\widetilde{E}(\omega_{fn_{r}}+\Delta\Omega) \widetilde{E}^{*}(\omega_{fn_{r}}) \right. \hspace{5cm} \nonumber \\  & & \left. 
         - \wp \int_{-\infty}^{+\infty} d\delta' \frac{1}{\delta'} \widetilde{E}(\omega_{fn_{r}}+\Delta\Omega-\delta')
                                                                   \widetilde{E}^{*}(\omega_{fn_{r}}-\delta') \right]. \hspace{0.7cm}  
\label{eq4:amp-4th-order} \vspace{-3cm}
\end{eqnarray}
These equations reflect the fact that $A_{f}^{(4)}$ interferes all
the four-photon pathways from $\left|g\right>$ to $\left|f\right>$
of any combination of three absorbed photons and one emitted photon.

Figure~\ref{fig_1} shows several representative four-photon pathways.
Each pathway can be divided into two parts: a non-resonant
absorption of two photons with a frequency sum of $\Omega =
\omega_{fg} - \delta$ and a Raman transition of two photons with a
frequency difference of $\Delta\Omega = \delta$.
Their border line is detuned by $\delta$ from either
$\left|f\right>$ or $\left|g\right>$ according to whether,
respectively, the two-photon absorption part or the Raman part comes
first (see Fig.~\ref{fig_1}).
The on-resonant ($\delta = 0$) and near-resonant ($\delta \ne 0$)
pathways are interfered separately in Eq.~(\ref{eq1:amp-4th-order}).
The Cauchy's principle value operator $\wp$ excludes the on-resonant pathways from the second term. 
The integration over the pathways is
expressed using 
two parameterized amplitudes, $A^{(2)}(\Omega)$ and $A^{(R)}(\Delta\Omega)$,
that originate 
from the different parts of the four-photon pathways.
$A^{(2)}(\Omega)$ 
interferes all the two-photon absorption pathways of 
frequency $\Omega$.
$A^{(R)}(\Delta\Omega)$ 
interferes all the Raman pathways of frequency $\Delta\Omega$ and
includes two components. The first is
$A^{(\scriptsize{\textrm{non-res}}R)}(\Delta\Omega)$ interfering all
the Raman pathways that are non-resonant,
with $\mu^{2}_{gg}$ and $\mu^{2}_{ff}$ being the 
$\left|g\right>$-$\left|g\right>$ and $\left|f\right>$-$\left|f\right>$
effective non-resonant Raman couplings due to 
the non-resonantly coupled states $\left|n\right>$.
The second is $A^{(\scriptsize{\textrm{res}}R)}(\Delta\Omega)$
interfering all the Raman pathways that are resonance-mediated via
$\left|n_{r}\right>$, with $\mu_{fn_{r}}$ being the
$\left|f\right>$-$\left|n_{r}\right>$ dipole matrix element.

For a given pulse shape $\widetilde{E}(\omega)$,
a non-zero $A_{f}^{(2)}$ is proportional to the maximal spectral intensity $I_{0} = |E_{0}|^{2}$
while a non-zero $A_{f}^{(4)}$ is proportional to $I_{0}^{2} = |E_{0}|^{4}$.
Different 
intensity $I_{0}$ corresponds to a different temporal peak intensity of the TL pulse ($I_{\scriptsize{\textrm{TL}}}$).
In this work, for a set of intensities $I_{0}$, the
$\left|f\right>$'s final population $P_{f} =
\left|A_{f}\right|^{2}$, which reflects the degree of two-photon
absorption, is controlled via the pulse shape
$\widetilde{E}(\omega)$.
The 
spectral phases $\Phi(\omega)$ are the corresponding control knobs,
with the spectrum $\left|\widetilde{E}(\omega)\right|$ unchanged.
The latter is chosen such that  
the central frequency $\omega_{0}$ is detuned from $\omega_{fg}/2$
and $\left|\widetilde{E}(\omega_{fg}/2)\right|^{2}$$\approx$0.5.
As discussed below, this corresponds to the interesting case of
having $A_{f}^{(4)}$ negligible relative to $A_{f}^{(2)}$ in the
weak-field limit and comparable to $A_{f}^{(2)}$ in the upper
intermediate-field limit.
Their interplay enables the absorption enhancement beyond the TL
absorption.

The Na system \cite{NIST} includes the $3s$ ground state as
$\left|g\right>$, the $4s$ state as $\left|f\right>$, the manifold
of $p$ states as the $\left|n\right>$ manifold, and the $7p$ state
as $\left|n_{r}\right>$.
The transition frequencies $\omega_{fg} \equiv \omega_{4s,3s}$ and $\omega_{fn_{r}} \equiv \omega_{7p,4s}$
correspond, respectively, to two 777-nm photons and one 781.2-nm~photon.
The sodium is irradiated with phase-shaped linearly-polarized femtosecond
pulses having an 
intensity spectrum centered around 779.5~nm with 5~nm bandwidth ($\sim$180~fs TL duration).  
Experimentally, a sodium vapor in a heated cell is irradiated with
such pulses of variable energy after they undergo shaping in an
optical set-up with a pixelated liquid-crystal spatial light phase
modulator \cite{pulse_shaping}.
Upon focusing, the TL temporal peak intensity at the peak of the
spatial beam profile
$I_{\scriptsize{\textrm{TL}}}^{\scriptsize{\textrm{(profile-peak)}}}$
ranges from $10^{9}$ to 5$\times$10$^{10}$~W/cm$^2$.
Following the interaction with a pulse, the 
population excited to
the $4s$ state undergoes cascaded 
decay to the $3s$ state via the $3p$ state. The corresponding
$3p$-$3s$ fluorescence serves as the relative measure for the final
$4s$ population $P_f \equiv P_{4s}$. It is measured using a
spectrometer coupled to a time-gated camera system.
The measured signal results from an integration over the spatial beam profile. %

\begin{figure} [htbp]
\includegraphics[scale=0.5]{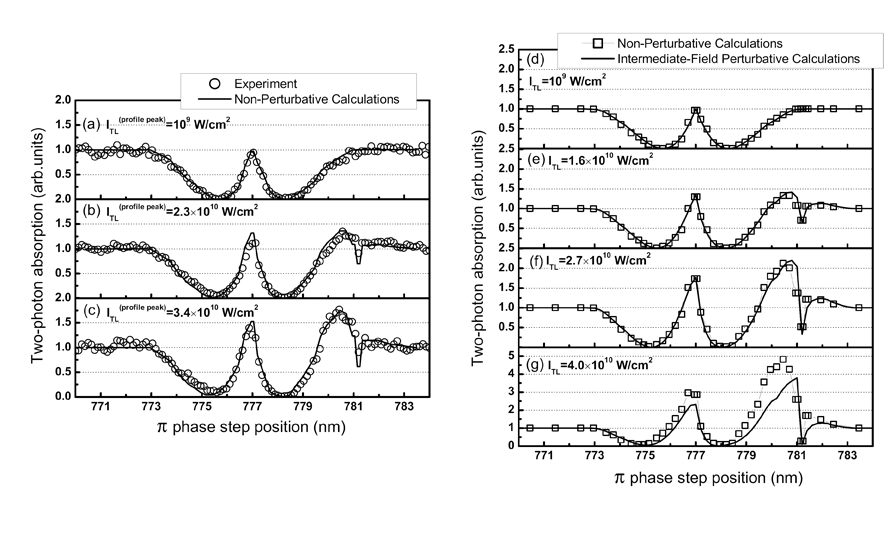} %
\caption{Results for the Na two-photon absorption induced by the
shaped pulses with a $\pi$ spectral phase step as a function of the
step position.
Panels (a)-(c): Experimental (circles) and non-perturbative theoretical (solid lines) results (spatially integrated) for different
pulse energies corresponding to a TL intensity at the peak of the spatial beam
profile $I_{\scriptsize{\textrm{TL}}}^{\scriptsize{\textrm{(profile-peak)}}}$
of (a) 10$^{9}$,  
(b) 2.3$\times$10$^{10}$, and (c) 3.4$\times$10$^{10}$~W/cm$^{2}$.
Panels (d)-(g): Non-perturbative (squares) and perturbative (solid lines) theoretical results for different
(single-valued) spectral intensities $I_{0}$ (see text) corresponding to a TL intensity $I_{\scriptsize{\textrm{TL}}}$ of
(d) 10$^{9}$, (e) 1.6$\times$10$^{10}$,
(f) 2.7$\times$10$^{10}$, and (g) 4$\times$10$^{10}$~W/cm$^{2}$.
The perturbative calculations include 2$^{nd}$ and 4$^{th}$ orders.
}
\label{fig_2}
\end{figure}

We validate our intermediate-field frequency-domain perturbative
description by comparing corresponding results to exact
non-perturbative results obtained by the numerical propagation of
the time-dependent Schr{\"o}dinger equation (using the Runga-Kutta
method). The considered manifold of $p$-states is from $3p$ to $8p$
\cite{NIST}. The validity of the non-perturbative calculations is
confirmed first by a comparison to experiment.
The test case is the set of shaped pulses having a $\pi$-step spectral phase pattern.
Each such pattern is characterized by the step position $\omega_{step}$,
with $\Phi(\omega \le \omega_{step}) = -\pi/2$ and $\Phi(\omega > \omega_{step}) = \pi/2$.
Figures~\ref{fig_2}(a)-(c) show examples of experimental (circles)
and non-perturbative theoretical (solid lines) results of $P_{4s}$
as a function of $\omega_{step}$ for different pulse energies, i.e.,
different~$I_{\scriptsize{\textrm{TL}}}^{\scriptsize{\textrm{(profile-peak)}}}$.
Each of the traces is normalized by $P_{4s}$ of the TL excitation.
The weak-field trace \cite{silberberg_2ph_nonres1_2} is shown in
Fig.~\ref{fig_2}(a).
The theoretical traces account for the experimental integration over the
beam profile: Each of them results from a set of 
calculations conducted each with a different (single) value of $I_{0}$.
As can be seen, the agreement between the experimental and
non-perturbative theoretical results is excellent, confirming these
calculations' accuracy.
Figs.~\ref{fig_2}(d)-(g) compare,
for different (single-valued) intensities $I_{0}$,
the theoretical non-perturbative results (squares) with
perturbative results 
calculated numerically using Eqs.~(\ref{eq:total-amp})-(\ref{eq4:amp-4th-order}) (solid lines).
Shown are 
examples out of the full set of results.
As can be seen, the perturbative results 
reproduce the non-perturbative ones up to the $I_{0}$ corresponding to a
TL peak intensity of $I_{\scriptsize{\textrm{TL}}}$$=$2.7$\times$10$^{10}$~W/cm$^{2}$ [Fig.~\ref{fig_2}(f)].
This is the intensity limit of the intermediate-field regime for the present Na excitation,
up to which no perturbative order beyond the 4$^{th}$ one is needed to be included.
The weak-field regime of $A_{4s}$$\approx$$A_{4s}^{(2)}$ extends here up 
to $I_{\scriptsize{\textrm{TL}}}$$\approx$10$^{9}$~W/cm$^{2}$ [Fig.~\ref{fig_2}(d)].

As shown in Figs.~\ref{fig_2}(d)-(f), the $\pi$-trace shape
significantly changes as $I_{0}$ increases.
The prominent feature reflecting the deviation from the weak-field regime is the 
absorption enhancement
beyond the TL absorption when $\omega_{step}$ is   
around $\omega_{4s,3s}/2$ (777~nm) or around the central spectral frequency $\omega_{0}$ (780~nm).
The enhancement increases as $I_{0}$ increases.
Here, using the intermediate-field description, we analyze      
the enhancement mechanism when $\omega_{step} = \omega_{4s,3s}/2$.
In the weak-field regime [Fig.~\ref{fig_2}(d)]
the corresponding 
absorption is equal to the (maximal)
TL absorption,
since such a phase pattern also leads to fully constructive
interferences within $A_{4s}^{(2)}$ \cite{silberberg_2ph_nonres1_2}.
In the present intermediate-field limit 
[Fig.~\ref{fig_2}(f)] it is about twice the TL absorption:
With the $I_{0}$ of $I_{\scriptsize{\textrm{TL}}}$$=$2.7$\times$10$^{10}$~W/cm$^{2}$
the calculated values are
$\{ A_{4s}^{(2)} = 0.35, A_{4s}^{(4)} = -0.15-0.10i : P_{4s} = 0.05 \}$ for the TL pulse and           
$\{ A_{4s}^{(2)} = 0.35, A_{4s}^{(4)} = -0.09-0.14i : P_{4s} = 0.09 \}$ %
for the pulse of $\omega_{step} = \omega_{4s,3s}/2$.
So, as seen from these values, the lower TL absorption 
results mainly from a stronger attenuation of the equal real 
$A_{4s}^{(2)}$ by $\Re[A_{4s}^{(4)}]$. 

The difference in $\Re[A_{4s}^{(4)}]$ for these two pulses originates from
the $\wp$-integral in Eq.~(\ref{eq1:amp-4th-order}), which 
interferes all the near-resonant four-photon pathways of $\delta \ne 0$ 
with the domination of small $|\delta|$ (due to the $1/\delta$ weighting).
Actually, it predominantly originates from the integrand factor
$A^{(2)}(\omega_{4s,3s}-\delta)$ interfering all the two-photon
absorption parts (of these four-photon pathways) of transition
frequency $\omega_{4s,3s} - \delta$. For the TL pulse, due to fully
constructive interferences,
$A^{(2)}(\omega_{4s,3s} - \delta)$ is maximized (real and positive value) for any $\delta$.         
So, with a detuned spectrum of $\omega_{0} \ne \omega_{4s,3s}/2$, it
changes monotonically with $\delta$ around~$\delta = 0$.
$A^{(2)}(\omega_{4s,3s} - |\delta|) > A^{(2)}(\omega_{4s,3s} +
|\delta|)$ with a red spectral detuning, and vice versa with a blue
detuning.
On the other hand, for the shaped pulse of
$\omega_{step}$=$\omega_{4s,3s}/2$,
$A^{(2)}(\omega_{4s,3s} - \delta)$ is maximized only for $\delta = 0$  
and gradually reduces for small non-zero $|\delta|$ 
with comparable magnitude for $\pm|\delta|$ (true for both red and
blue spectral detuning). Consequently, considering also the
$1/\delta$ integrand factor, the $\wp$-integral results in a higher
value of $\left|\Re[A_{4s}^{(4)}]\right|$ for the TL pulse.
For both pulses, the sign of $\Re[A_{4s}^{(4)}]$ relative to $A_{4s}^{(2)}$ is
determined by the pulse spectrum and by the Raman couplings 
of the four-photon pathways
($\mu^{2}_{3s,3s}$, $\mu^{2}_{4s,4s}$ and $\left|\mu_{4s,7p}\right|^{2}$).
Here, the present red-detuned spectrum ($\omega_{0}$ of 779.5 nm)
leads to $A_{4s}^{(2)}$ and $\Re[A_{4s}^{(4)}]$ that are of opposite
signs. Shifting the spectrum to be comparably blue detuned from
$\omega_{4s,3s}/2$ ($\omega_{0}$ of 774.5 nm) changes their signs to
be the same.
Then, the intermediate-field absorption of the TL pulse is 
the higher one. A non-detuned spectrum ($\omega_{0}$ of 777 nm)
leads to a negligible $\Re[A_{4s}^{(4)}]$ and thus to an equal
absorption for both pulses, in agreement
with the strong-field time-domain study by Dudovich {\it et al.} \cite{silberberg-2ph-strong-field}.  
This spectral dependence is the subject of future publication.
For the current red-detuning, 
the above explanation also allows to understand why the enhancement over the TL absorption increases as $I_{0}$ increases:
Since $\left|\Re[A_{4s}^{(4)}]\right| / A_{4s}^{(2)} = K I_{0}$ with
higher value of $K$ for the TL pulse as compared to the shaped pulse
of $\omega_{step}$=$\omega_{4s,3s}/2$,
a given increase of $I_{0}$ results in 
a higher increase in the attenuation of
$A_{4s}^{(2)}$ by $A_{4s}^{(4)}$ for the TL pulse.
%

\begin{figure} [htbp]
\includegraphics[scale=0.5]{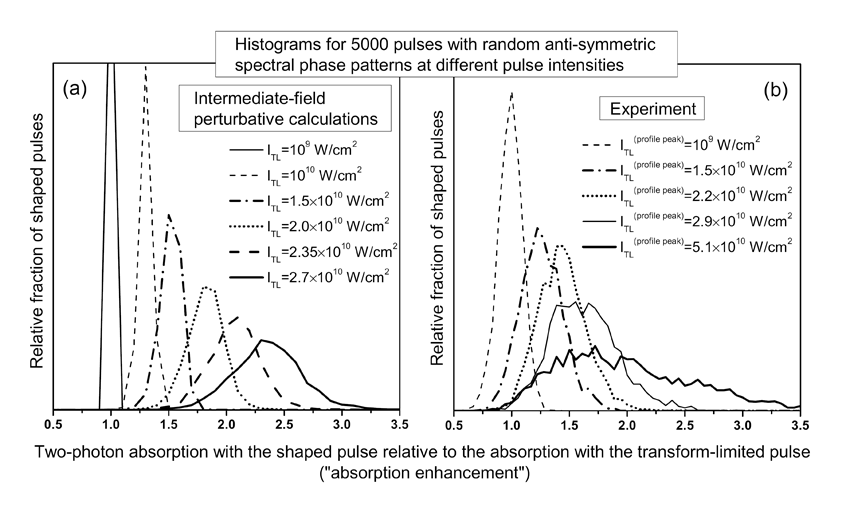} %
\caption{Histograms of the Na two-photon absorption for a set of
5000 different pulses
with random spectral phase patterns that are all anti-symmetric around $\omega_{4s,3s}/2$.  
They show the fraction of pulses inducing different absorption enhancement over the TL pulse.
(a) Theoretical perturbative results (of 2$^{nd}$ and 4$^{th}$
orders) for different (single-valued) spectral intensities $I_{0}$
corresponding to different TL intensities
$I_{\scriptsize{\textrm{TL}}}$.
(b) Experimental results (spatially integrated) for different pulse energies corresponding to different 
TL intensities at the peak of the beam profile
$I_{\scriptsize{\textrm{TL}}}^{\scriptsize{\textrm{(profile-peak)}}}$.
} \label{fig_3}
\end{figure}

Similar to the $\pi$-step phase pattern of $\omega_{step}$=$\omega_{4s,3s}/2$,
all the spectral phase patterns that are anti-symmetric around 
$\omega_{4s,3s}/2$, i.e., $\Phi(\omega) = - \Phi(\omega_{4s,3s}-\omega)$, %
lead to the maximal weak-field two-photon absorption
for, and only for, a 
transition frequency of $\omega_{4s,3s}$ \cite{silberberg_2ph_nonres1_2}.
So, based on the above analysis, in the intermediate-field regime
the corresponding shaped pulses are expected to enhance the two-photon absorption (to the $4s$ state)
beyond the TL absorption.
The mechanism is the 
enhanced attenuation of $A_{4s}^{(2)}$ by $A_{4s}^{(4)}$ for the TL pulse. 
Figure~\ref{fig_3} confirms that indeed this is the case.     
It shows theoretical [Fig.~\ref{fig_3}(a)] and experimental [Fig.~\ref{fig_3}(b)] results
for 5000 different pulse shapes with such anti-symmetric phase patterns that we have randomized.
The results are presented as histograms (distributions) showing
the fraction of pulses inducing different values of $P_{4s} / P_{4s}^{(\scriptsize{\textrm{TL}})}$,
i.e., different absorption enhancement over to the TL excitation.
The theoretical results are perturbative
and have been calculated numerically using Eqs.~(\ref{eq:total-amp})-(\ref{eq4:amp-4th-order}), with
each histogram corresponding to a different (single-valued) intensity $I_{0}$.
The (spatially-integrated)
experimental histograms correspond to different pulse energies.
The weak-field histogram is located at the value~of~1. 

As can be seen, the distribution shifts to higher enhancement values
and gets broader as $I_{0}$ increases. Both effects originate from
the $I_{0}$-dependence of $\left|\Re[A_{4s}^{(4)}]\right| /
A_{4s}^{(2)} = K I_{0}$, with $K$ depending only on the pulse shape.
Similar to the 
$\omega_{step}$=$\omega_{4s,3s}/2$ case,
the shift to higher enhancement values originates from the higher $K$ value
for the TL pulse
as compared to the shaped pulses with the anti-symmetric phase patterns.
The distribution broadening originates from the variation in
the $K$ value from one pulse shape to the other.
The experimental histograms are generally broader than the theoretical ones due to the integration over  
the spatial beam profile 
and due to the experimental noise.
The experimental signal-to-noise ratio is directly reflected in the width of the weak-field histogram
[Fig.~\ref{fig_3}(b)-dashed line].
Our most-enhancing random anti-symmetric 
pattern is calculated at $I_{0}$ of
$I_{\scriptsize{\textrm{TL}}}$=2.7$\times$10$^{10}$~W/cm$^{2}$
to induce $P_{4s}$=0.21, corresponding 
to 4.2 enhancement over the TL absorption. 
Last, it worth mentioning that
excluding the resonantly-coupled $7p$ state from our calculations 
generally leads to about 25\% reduction in the enhancement factors over the TL absorption
at $I_{\scriptsize{\textrm{TL}}}$=2.7$\times$10$^{10}$~W/cm$^{2}$.

In summary, we have extended the powerful frequency-domain analysis of femtosecond 
two-photon absorption to the intermediate-field regime of 2$^{nd}$ and 4$^{th}$ perturbative orders.
This opens the door for rational femtosecond coherent control 
beyond the weak-field regime, in a regime of considerable absorption yields. 
The extended frequency-domain description has enabled us to identify a broad family of shaped pulses that 
enhance the absorption beyond the TL absorption. 
The mechanism is the coherent interplay between the different perturbative orders.
We expect this work to serve as a basis for 
intermediate-field extensions to molecules 
and other 
multiphoton processes as well as for
devising effective control strategies 
when several multiphoton channels are 
involved.
The results 
are also of applicative importance for nonlinear spectroscopy and microscopy. 

\newpage

\begin{thebibliography}{9999}
%
\bibitem{tannor_kosloff_rice_coh_cont} D. J. Tannor, R. Kosloff,
and S. A. Rice, J. Chem. Phys. $\boldsymbol{85}$, 5805 (1986).

\bibitem{shapiro_brumer_coh_cont_book} M. Shapiro and P. Brumer,
{\it Principles of the quantum control of molecular processes}
(Wiley, New Jersey, 2003).

\bibitem{warren_rabitz_coh_cont} W. S. Warren, H. Rabitz, and D. Mahleh,
Science $\boldsymbol{259}$, 1581 (1993).

\bibitem{rabitz_vivie_motzkus_kompa_coh_cont} H. Rabitz, R. de Vivie-Riedle,
M. Motzkus, and K. Kompa, Science $\boldsymbol{288}$, 824 (2000).


\bibitem{dantus_exp_review1_2} M. Dantus and V. V. Lozovoy,
Chem. Rev. $\boldsymbol{104}$, 1813 (2004);
ChemPhysChem $\boldsymbol{6}$, 1970 (2005).   


\bibitem{silberberg_2ph_nonres1_2} D. Meshulach and Y. Silberberg,
Nature (London) $\boldsymbol{396}$, 239 (1998);
Phys. Rev. A $\boldsymbol{60}$, 1287 (1999).


\bibitem{dantus_2ph_nonres_molec1_2} K. A. Walowicz {\it et al.},
J. Phys. Chem. A $\boldsymbol{106}$, 9369 (2002);
V. V. Lozovoy {\it et al.}, J. Chem. Phys. $\boldsymbol{118}$, 3187 (2003).


\bibitem{baumert_2ph_nonres} A. Pr$\ddot{a}$kelt {\it et al.},
Phys. Rev. A $\boldsymbol{70}$, 063407 (2004).   

\bibitem{silberberg_2ph_1plus1} N. Dudovich {\it et al.},
Phys. Rev. Lett. $\boldsymbol{86}$, 47 (2001).


\bibitem{girard_2ph_1plus1} B. Chatel, J. Degert, and B. Girard,
Phys. Rev. A $\boldsymbol{70}$, 053414 (2004).

\bibitem{becker_2ph_1plus1_theo}
P. Panek and A. Becker, Phys. Rev. A $\boldsymbol{74}$, 023408 (2006).

\bibitem{silberberg_antiStokes_Raman_spect}
D. Oron {\it et al.}, Phys. Rev. A $\boldsymbol{65}$, 043408 (2002);
N. Dudovich, D. Oron, and Y. Silberberg, Nature (London) $\boldsymbol{418}$, 512 (2002);


\bibitem{leone_res_nonres_raman_control}
H. U. Stauffer {\it et al.}, J. Chem. Phys. $\boldsymbol{116}$, 946 (2002);
X. Dai, E. W. Lerch, and S. R. Leone, Phys. Rev. A $\boldsymbol{73}$, 023404 (2006).

\bibitem{amitay_3ph_2plus1} A. Gandman, L. Chuntonov, L. Rybak, and Z. Amitay,
Phys. Rev. A $\boldsymbol{75}$, 031401 (R) (2007).

\bibitem{gersh_murnane_kapteyn_Raman_spect}
E. Gershgoren {\it et al.}, Opt. Lett. $\boldsymbol{28}$, 361 (2003).

\bibitem{silberberg-2ph-strong-field} N. Dudovich {\it et al.},
Phys. Rev. Lett. $\boldsymbol{94}$, 083002 (2005).

\bibitem{weinacht-2ph-strong-theo-exp} C. Trallero-Herrero {\it et al.},
Phys. Rev. A. $\boldsymbol{71}$, 013423 (2005);
Phys. Rev. Lett. $\boldsymbol{96}$, 063603 (2006).


\bibitem{wollenhaupt-baumert1_2} M. Wollenhaupt {\it et al.}, Phys. Rev. A $\boldsymbol{68}$, 015401 (2003);
Phys. Rev. A $\boldsymbol{73}$, 063409 (2006);


\bibitem{hosseini_theo_multiphoton_strong}
S. A. Hosseini and D. Goswami, Phys. Rev. A $\boldsymbol{64}$, 033410 (2001).



\bibitem{rabitz_feedback_learning_idea} R. Judson and H. Rabitz, Phys.
Rev. Lett. $\boldsymbol{68}$, 1500 (1992).

\bibitem{pulse_shaping} A. M. Weiner, Rev. Sci. Inst. $\boldsymbol{71}$, 1929 (2000).

\bibitem{NIST}
NIST Atomic Spectra Database (National Institute of Standards and
Technology, Gaithersburg, MD).


%
\end{thebibliography}
\end{document}